# Distribution of the magnetization reversal duration in sub-ns spin-transfer switching

T. Devolder[a], C. Chappert[a], J.A. Katine[c], M.J. Carey[c] and K. Ito[b]

[a] Institut d'Electronique Fondamentale, CNRS UMR 8622, Université Paris Sud, Bât. 220, 91405 Orsay, France.

[b] Hitachi Cambridge Laboratory, Hitachi Europe, Ltd., Cavendish Laboratory, Madingley Road, Cambridge CB3 0HE, UK

[c] Hitachi GST, San Jose Research Center, 650 Harry Road, San Jose, California 95120, USA

_Abstract_: We study the distribution of switching times in spin-transfer switching induced by sub-ns current pulses in pillar-shaped spin-valves. The pulse durations leading to switching follow a comb-like distribution, multiply-peaked at a few most probable, regularly spaced switching durations. These durations reflect the precessional nature of the switching, which occurs through a fluctuating integer number of precession cycles. This can be modeled considering the thermal variance of the initial magnetization orientations and the occurrence of vanishing total torque in the possible magnetization trajectories. Biasing the spin-valve with a hard axis field prevents some of these occurrences, and can provide an almost perfect reproducibility of the switching duration.

The spin-transfer effect [1] is the exchange of angular momentum between a spin-polarized electrical current and the magnetization of a nanomagnet. The spin-transfer results in torques that can be used to manipulate magnetic configurations with a sole current. When a spin-transfer torque (STT) is used in a magneto-resistive system, the current can play two roles, since the electrical resistance can be used to probe the configuration that the current manipulates. In recent years, STT has been achieved in a variety of systems, leading to new phenomena such as non-ohmic behavior in metallic multilayers [2], displacement of domain walls [3], generation of spin waves [4], and pumping of small [5] or large [6] amplitude steady state magnetization precessions.

STT can also simply switch the magnetization of a uniaxial nanomagnet [7], which is considered as a promising route for memory applications [8], since this type of switching has proven deep sub-ns potential [9, 10]. However, previous investigations have concluded that the reversal speed in the sub-ns regime has insufficient reproducibility. This has first been interpreted qualitatively as resulting from classical thermal fluctuations [11], but reliable predictions are not yet available.

In this letter, we show experimentally that the sub-ns pulse durations leading to successful switching events are *discrete* quantities reflecting the precessional nature of magnetization dynamics, and the topological peculiarities in the set of possible magnetization trajectories. This tendency towards quantization of the switching times can be manipulated using a hard axis field to lift the near degeneracy between magnetic trajectories. We discuss these findings by taking into account the precessional dynamics, the STT and the thermal effects.

Our devices are spin valves of composition PtMn17.5/CoFe1.8/Ru8/CoFe2/Cu3.5/CoFe1/NiFe1.8 (thickness in nm), etched into elongated hexagons, whose major axis is parallel to the PtMn exchange pinning direction. We have investigated two sizes: 75×150 nm² (category A) and 75×113 nm² (category B); they yield similar results. The devices are similar to those used in ref. [12], except that here they are inserted in a high frequency layout. Their properties are described elsewhere [13]. For category B, the mean quasi-static Parallel state to AntiParallel state (P→AP) and AP→P switching currents are -3 mA (i.e. -1.1 ×10$^8$ A/cm²) and 1.1 mA (i.e. 5.3×10$^7$ A/cm²), respectively.

The measurement procedure for AP to P switching is the following. The sample is first prepared in the AP state by $I_{DC}$. $R_{AP}$ is measured at remanence. The current pulse $I_{pulse}$ is then applied and the resistance $R'$ is determined after relaxation. A negative $I_{DC}$ is then applied to ensure returns to the P state, and $R_P$ is measured. The ratios $(R'-R_P)/(R_{AP}-R_P)$ are very near 1 or 0, indicating that the reversal is either complete or non-existant. Each ratio is thus used to decide whether switching has occurred for a given current pulse duration and amplitude. The procedure is repeated 1000 times for each pulse amplitude/duration. To estimate the switching probability $p$ versus $I_{pulse}$ and $\tau_{pulse}$, we measure $n$ successful switching events out of $N=1000$ trials and say that $p \approx n/N$. Our finite number of trials results in a random gaussian error $\Delta n/\langle n \rangle = \sqrt{p(1-p)/N}$. This error is at worst 1.6% when $p=50\%$. Note that our procedure, in contrast to time-domain averaging of the magnetization response [14, 15], is able to detect rare events. We shall see that this is important since some specific initial conditions lead to a quasi-divergence of the switching time.

In Figure 1A, we show the switching probability for the AP→P transitions for a sample of category B and for pulse durations $\tau_{pulse}$ from 100 ps to 10 ns. Our results follow the rule of thumb that the switching requires a pulse duration $\tau_{pulse}$ that scales with the inverse of the overdrive current $\Delta = (J - J_{C0})/J_{C0} > 0$, where $J_{C0} \approx \alpha(\mu_0 M_S^2 t |e|)/(2 p \hbar) \approx 10^7$ A/cm² is the zero-temperature switching current, with $t$ the free layer thickness, and $p$ the effective spin polarization. We write the applied current density as $J_{applied} = (\Delta + 1) J_{C0}$. The surface of the switching/no switching boundary indicates that the switching duration has a dispersion of about ±30%.

In Fig. 1B, we have zoomed on some horizontal cuts of Fig. 1A: we set $J_{applied}$, vary the pulse duration $\tau_{pulse}$ by increments of 10 ps, and evaluate the resulting switching probability. The switching probability increases with $\tau_{pulse}$, but this increase is not regular: flat plateaus alternate with rounded steps (Fig.1B, bottom curves). This behavior has been observed systematically for AP to P switching for $0.1 < \tau_{pulse} < 1.2$ns. The steps are better revealed when looking at the differential switching probability density (Fig. 1B, top curves). The latter describes the probability that the reversal is induced between $t$ and $t+dt$. It has a comb-like structure, with most often two peaks at the most probable switching times. The same trends have been observed for the reverse (i.e. P→AP) transition, however with generally a fainter step-to-plateau contrast.

When we vary the current, we observe correlations between the step positions, indicating that they may reflect some periodicity in the magnetization reversal paths. We have thus gathered in Fig. 1D the most probable switching durations (i.e. the step positions) versus $I_{pulse}$ for the AP→P transition in samples B. The peak positions are labeled with symbols/colors according to their index. The shortest switching durations (black squares) are grouped between pulse durations of 110 and 140 ps. They are observable only at currents higher than 9 mA. The second probable switching durations (red circles) are grouped in the interval between 320 and 400 ps; they were identified for all the studied applied currents (Fig. 1B). Most of the third probable switching durations (blue triangles) arise between 560 and 630 ps.

Finally, we have performed additional measurements after applying a constant field $H_y$ along the hard axis. A representative result is reported in Fig. 1C; we display the switching probability distribution for the P to AP transition with a constant overdrive (Δ=6) and a *variable hard axis applied field*. We can notice that for $H_y$=0, there is a single plateau at around $\tau_p$=150 ps, and switching probability curve is not that rich. When the field is raised to 2.8 mT, (in this case, this corresponds to $H_k/4$), there appears a highly probable switching duration at 290 ps. When the field is further increased to 4.2 mT, the probability of switching in 290 ps is further reinforced, while the step at 150 ps disappears. More generally, we have observed that for P to AP as well as for AP to P transitions, adding a hard axis field reinforces or reveals some steps, while it make some other disappear.

In order to understand the origin of this comb-like distribution of switching durations, we have performed extensive simulations. For this, we model the behavior of our well characterized [13] category B of samples using a thin macrospin lying in the *(xy)* plane, having a magnetization $M_S$=6.76×10$^5$ A/m ($\mu_0 M_S$=0.85 T), a thickness t=2.8 nm, a uniaxial anisotropy of $\mu_0 H_k$=20 mT along an easy axis *(x)* and a Gilbert damping parameter α=0.02. The current carries a spin polarisation p=0.27 along *(-x)*. We use a sinusoidal angular dependance of the STT, and the standard Landau-Lifchitz-Gilbert equation. Before the square current pulse is applied, the normalized magnetization is assumed to be $\mathbf{m_0}$ ={$m_{y0}$, $m_{y0}$, 0} i.e. in the film plane, near its equilibrium position.

Let us first consider the switching when there is no applied field ($H_y$=0). Representative reversal trajectories are displayed in Fig. 2A and 2B for $m_{y0}$=0.128 and two very near overdrive parameters of Δ=3.02 (black) and Δ=3.05 (red). The magnetization undergoes first an elliptical precession around its easy axis, with a growing precession amplitude. Depending on $\mathbf{m_0}$ and $J_{appl}$, the reversal proceeds through some finite Number of Half Precession (*NHP*) cycles before magnetization overcomes the hard axis. We define *NHP* as the sum of the number of maxima and minima in the trace of $m_y(t)$, including that occurring when $m_x$=0. This definition of *NHP* is straightforward when counting the number of turn in Fig. 2B before the cusp in the {$m_y$, $m_z$} trajectories. *NHP* is odd (even) when the magnetization switching occurs by a clockwise (counterclockwise) rotation in the {$m_x$, $m_y$} plane (see insets in Fig. 1A and 1B). We write CW (CCW) for clockwise (respectively counterclockwise) rotations. After having overcome the hard axis, the magnetization finally relaxes to the reverse easy axis position ($m_x$=-1) following a heavily damped precessional trajectory.
While *NHP* is 3 for Δ=3.05, a marginally smaller overdrive Δ=3.02 leads to *NHP*=4. In between 3.02<Δ<3.05 , there is a remarkable current density $J_{bif}$. For $J_{appl}=J_{bif}$, the magnetization passes at a specific orientation (cross in Fig. 2A) with $m_x=0$ where the demagnetizing and the spin torques cancel each other; the magnetization feels a *vanishing total torque,* and a perturbation is needed to either switch immediately or perform another half precession cycle before indeed switching. At this specific magnetization orientation, the magnetization {$m_x$, $m_y$, $m_z$} follows $m_x=0$ and :

$$\text{Eq. 1} \quad \frac{J_{max\,Jitter}}{m_z m_y} = \frac{\mu_0 M_S^2 t |e|}{p \hbar}$$

The reason why we index this current density as the "maximum jitter" current density will appear clearly latter in the discussion. Note that since in practice the applied current satisfies $\Delta\alpha \ll 2$ , there exists two initial conditions leading to Eq. 1.

The happening of a vanishing total torque yields two important consequences for the switching duration $\tau_{m_x=0}$ and its repeatability. For stochastic initial conditions, the possibility of satisfying Eq. 1 may add an incremental *jitter* of exactly half a precession period to the overall switching time. This results in a step-like dependence of $\tau_{m_x=0}$ versus current at given $m_{y0}$ (see Fig. 2C) , or to a step-like dependence of $\tau_{m_x=0}$ versus $m_{y0}$ (see Fig. 2D) at given current. In addition, when approaching a vanishing total torque magnetization orientation (Eq. 1), the reversal time $\tau_{m_x=0}$ diverges (see Fig. 2C and 2D). In figure Fig. 3A, we report the switching times versus both the initial magnetization orientation $m_{y0}$ and the overdrive current, in zero field. The divergence of the switching time for initial magnetization along the easy axis ($m_{x0}=1$) appears as a horizontal line. The conditions for vanishing total torques appear as curved contours, separating switching regions with differing *NHP*.

Due to the finite temperature, each switching test encounters a different initial condition. We thus model the initial magnetization with an orientation randomly distributed in the sample plane, following Boltzmann statistics. This distribution (Fig. 2D) has a width of $m_{y0}^{rms} = \sqrt{(kT)/(\mu_0 H_k M_S V)}$ . This width is 0.13 at T=300K. It is sketched as the vertical segment in Fig. 3A.

With Δ=3, a majority of switching events requires *NHP* to be 4 or 5 while a few events will require more *NHP*. As a result, the probability of successful switching with a given overdrive shall increase step-like with the pulse duration, i.e. the switching durations should follow a comb-like distribution. Such behavior is calculated in Fig. 4A for overdrives ranging from 3 to 7. These calculated distributions of switching times (Fig. 4A) compares quite well with experiments (Fig. 1B). A quantitative agreement is even obtained between the experimental most probable switching times (Fig. 1D) and the step positions (dotted lines) in Fig. 4A. Note also that in the calculations, the most probable switching times slightly shift to small durations when the overdrive increases (see dotted lines in Fig. 4A).

We now simulate the effect of a hard axis field. The initial magnetization orientation is chosen near its equilibrium $<m_{y0}>= H_y/H_k$, with a variance still assumed to follow Boltzmann statistics. Once again, the current pulses primarily amplify the precession (not shown). However, as soon as the current is applied, the spin-torque gets finite with an out-of-the-film-plane component being $pJ\hbar m_{y0}/(\mu_0 M_S t |e|)$. Hence, in contrast to the zero field case, there is *no divergence* of the switching speed when the magnetization is exactly along its in-field equilibrium position (compare Fig. 3A and 3B). This removal of divergence is a clear benefit of applying a hard axis field.

However, there is another important change induced by the applied field. Without applied fields, switching by CW rotation (odd *NHP)* or CCW rotation (even *NHP)* takes place with comparable probabilities. This does not hold when $H_y \neq 0$: small overdrives only lead CCW rotations, and the reversal proceeds by passing near the hard direction favored by the field. This is illustrated in Fig.3 B, which summarizes the switching times versus $m_{y0}$ and $\Delta$, when a static hard axis field $H_y=0.25H_k$ is applied. The crossing of a vanishing total torque contours at overdrives $\Delta < 3$ corresponds to changes in *NHP by increments of two units* and the switching is always of CCW nature. Only at larger overdrives, some initial conditions $m_{y0}$ can lead to CW rotations, and the crossings of a vanishing total torque contours change the *NHP* by increments of one unit.

As done previously in zero applied field, we can calculate for $H_y \neq 0$ the switching probability versus pulse durations (Fig.4 B). At overdrives $\Delta \leq 4$, the various possible initial magnetization orientations can lead to several *NHP* values, such that the switching probability is multiply stepped versus the pulse duration. However, since the *NHP* can only take even values, the number of steps is typically twice less than in the zero-field case (Fig. 4A). This correlates well with our experimental results.

Interestingly, at overdrives $\Delta=5$ and 6, almost any initial magnetization within the Bolzmann distribution leads to a reversal taking *NHP*=2. The reversal duration is thus expected to be very reproducible from one switching event to the next (Fig. 4B); this is one of of central findings in this paper. Attempting to further accelerate the switching is increasing the overdrive current to 7 is detrimental to the reproducibility of the switching duration, because it introduces two drawbacks. It first triggers some faster reversal events requiring one less precession cycle. An overdrive current of 7 also generates a small number of very slow switching events, resulting from initial conditions matching a vanishing total torque criterion. This appears as a slow saturation of the switching probability above pulse durations of 300 ps.

In summary, we have studied sub-ns spin-transfer switching. The current pulse durations leading to switching follow a comb-like distribution. Modeling indicates that depending on the initial magnetization and the current amplitude, specific vanishing total torque magnetization position in the possible magnetization trajectories make the switching duration jitter by increments of the half precession period. The nature of these vanishing torque positions also implies that the reversal time diverges for a few initial conditions, a problem that is lifted as soon as a hard axis field is applied. At large overdrives, this hard axis applied field can even suppress the vanishing torque positions, allowing an almost perfect reproducibility of the switching time.

Figure captions

Figure 1: (color online) **(A)**: Experimental AP to P switching probability versus current pulse magnitude and duration for a sample of category B. The gray level scales with the switching probability. **(B)**: Horizontal cuts through Fig. 1A: AP to P switching probability versus current pulse duration for -8, -9, -11, -13 and -14 mA current pulses. Top curves: distribution of the switching times for -13 mA (blue) and for -8 mA (magenta) **(C)**: experimental P to AP switching probability for a sample of category A submitted to 32 mA and static hard axis fields. **(D)** : Applied current dependance of the most probable switching times for AP to P switching in a sample of type B. The horizontal lines are guides to the eyes.

Figure 2: (color online) **(A, B)**: Calculated magnetization trajectories after the application of two current steps of slightly different amplitude, in zero applied field and with an initial magnetization following $m_{y0}=0.128$, $m_{z0}=0$. Insets: in-plane projection of the magnetization trajectories, showing whether the switching happens by clockwise and counterclockwise rotation **(C)**: calculated switching times versus overdrive current with initial magnetization following $m_{y0}=0.128$, $m_{z0}=0$, and for $H_y=0$ or $\mu_0 H_y=0.128\mu_0 H_k=2.5$ mT. **(D)** : Calculated switching time versus initial magnetization orientation in zero applied field and in a current corresponding to an overdrive parameter of 3. Grey curve: Bolzmann distribution of the initial magnetization.

Figure 3: (color online) Calculated switching time versus initial magnetization orientation and overdrive current, in **(A)**: zero applied field. The vertical segment indicate the width of the thermal distribution of initial states. **(B)**: an applied field of $H_y=0.13H_k$, i.e. $\mu_0 H_y=2.5$ mT. The numbers superimposed on the graph indicate the Number of Half Precession cycles needed for magnetization reversal.

Figure 4: (color online) Calculated Switching probabilities versus pulsed current duration at temperatures for several overdrive currents. **(A)**: in zero applied field. The near vertical dotted lines correspond to the most probable switching times. **(B)**: in a hard axis applied field of $H_y=0.25H_k$.

Figure 1:

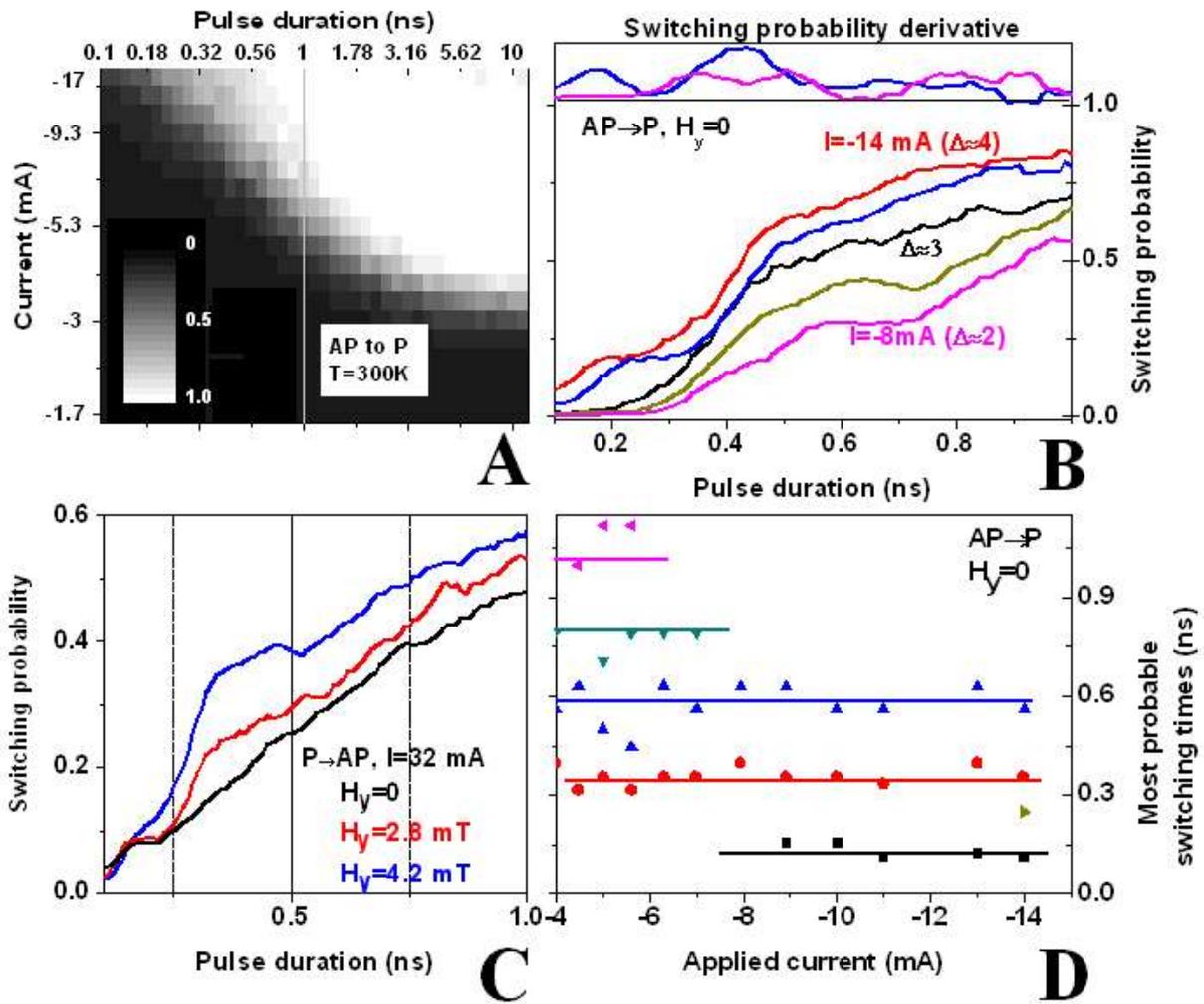

Figure 2:

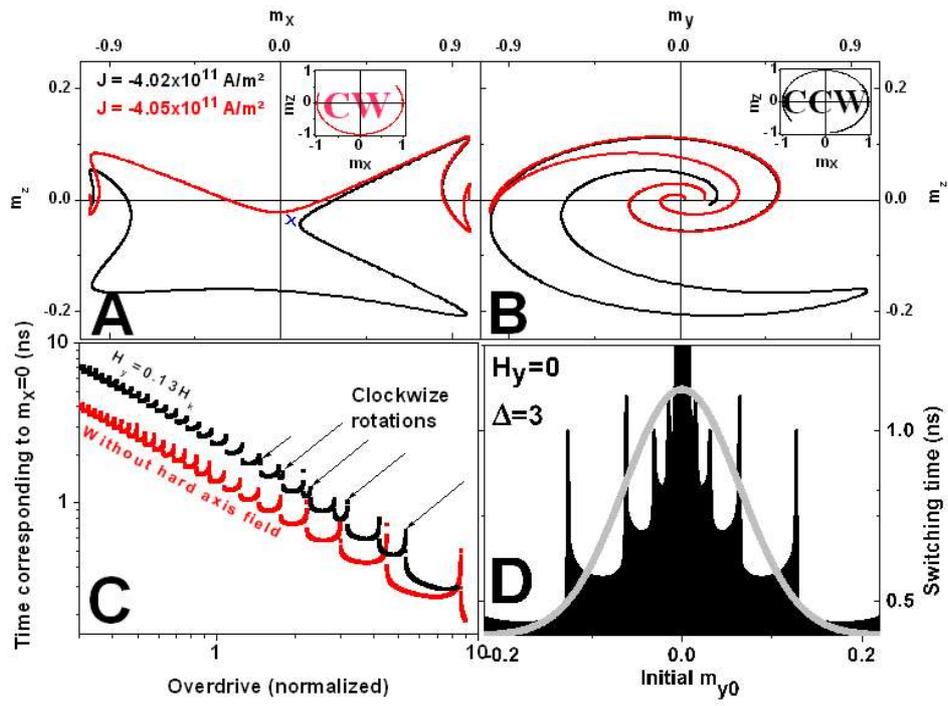

Figure 3:

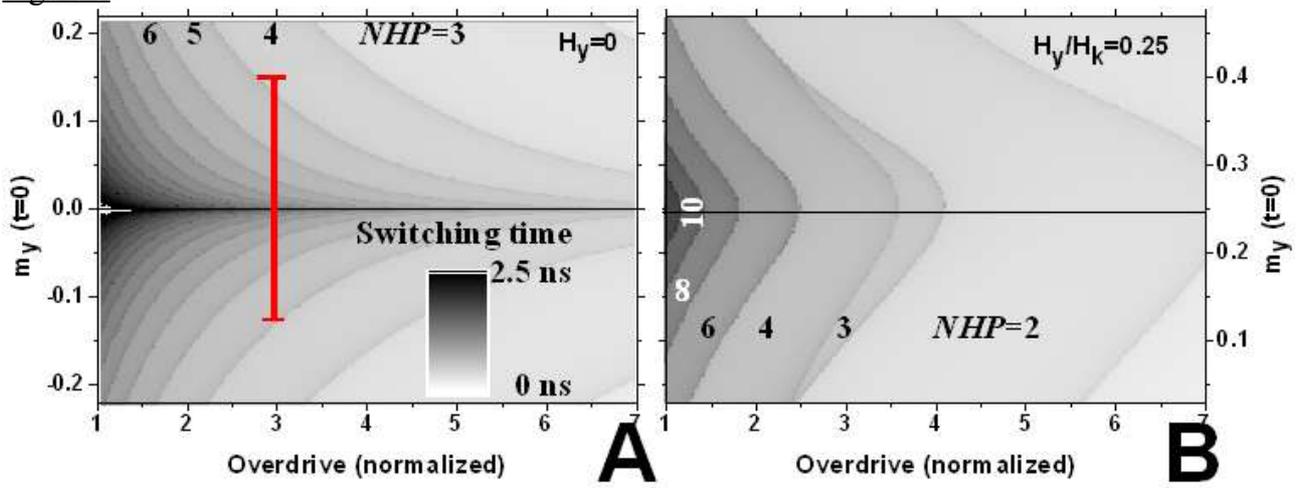

Figure 4:

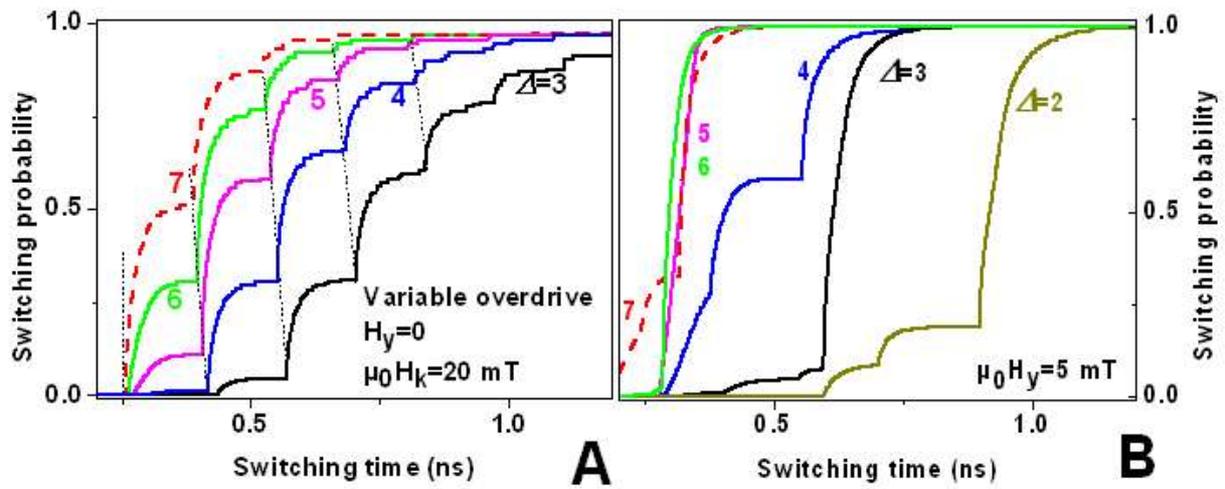

References:

[1] J. Slonczewski, *J. Magn. Magn. Mater.* **159**, 1 (1996).
[2] M. Tsoi et al. *Phys. Rev. Lett*. **80**, 4281 (1998)
[3] E. Saitoh, H. Miyajima, T. Yamaoka, G. Tatara, *Nature* **432**, 2003 (2004).
[4] B. Özyilmaz, A. D. Kent, J. Z. Sun, M. J. Rooks, and R. H. Koch, *Phys. Rev. Lett*. **93**, 176604 (2004) .
[5] T. Devolder, P. Crozat, C. Chappert, J. Miltat, A. Tulapurkar, Y. Suzuki, K. Yagami,  *Phys. Rev. B* **71**, 184401 (2005).
[6] S. I. Kiselev, J. C. Sankey, I. N. Krivorotov, N. C. Emley, R. J. Schoelkopf, R. A. Buhrman, D. C. Ralph, *Nature* **425**, 380 (2003)..
[7] J. A. Katine, F. J. Albert, and R. A. Buhrman, *Appl. Phys. Lett*. **77**, 3809 (2000); J. A. Katine et al., *Phys. Rev. Lett*. **84**, 3149 (2000).
[8]  A Novel Nonvolatile Memory with Spin Torque Transfer Magnetization Switching: Spin-RAM, M. Hosomi, et al. proceedings of IEDM conference (2005).
[9]   A. A. Tulapurkar, T. Devolder, K. Yagami, P. Crozat, C. Chappert, A. Fukushima, Y. Suzuki. *Appl. Phys. Lett.*  **85**(22) 5358 (2004).
[10] T. Devolder , C. Chappert, P. Crozat, A. Tulapurkar, Y. Suzuki, J. Miltat, K. Yagami, *Appl. Phys. Lett.*     **86**, 062505 (2005).
[11] T. Devolder, A. Tulapurkar, Y. Suzuki, C. Chappert, P. Crozat,  K. Yagami, *J. Appl. Phys.* 98 053904-1  *(*2005).
[12] D. Lacour, J. A. Katine, N. Smith, M. J. Carey, and J. R. Childress, *Appl. Phys. Lett.* **85**(20) 4681 (2004).
[13] T. Devolder, K. Ito, J. A. Katine, P. Crozat, J.-V. Kim and C. Chappert. *Appl. Phys. Lett.*  **88**, 152502 (2006).
[14] N.C. Emley, I. N. Krivorotov, O. Ozatay, A.G.F. Garcia, J.C. Sankey, D. Ralph, R.A. Buhrman, *Phys. Rev. Lett.* **96**(24) 247204 (2006).
[15] Y. Acremann, J.P. Strachan, V. Chembrolu, S.D. Andrews, T. Tyliszczak, J.A. Katine, M.J. Carey, B.M. Clemens, H.C. Siegmann, J. Stohr, *Phys. Rev. Lett.* **96**(21), 217202 (2006).